# New Approach to Designing Functional Materials for Stealth Technology: Radar Experiment with Bilayer Absorbers and Optimization of the Reflection Loss

*Jaume Calvo-de la Rosa\*, Aleix Bou-Comas, Joan Manel Hernàndez, Pilar Marín, Jose Maria Lopez-Villegas, Javier Tejada, and Eugene M. Chudnovsky*

Jaume Calvo-de la Rosa, Joan Manel Hernàndez, Javier Tejada

Departament de Física de la Matèria Condensada, Universitat de Barcelona, Martí i Franquès 1, 08028 Barcelona, Spain

Institut de Nanociència i Nanotecnologia (IN2UB), Universitat de Barcelona, 08028 Barcelona, Spain

E-mail: jaumecalvo@ub.edu

Aleix Bou-Comas

Graduate Program in Physics and Initiative for the Theoretical Sciences, Graduate Center, The City University of New York, New York, NY 10016, USA

Pilar Marín

Instituto de Magnetismo Aplicado (IMA-UCM-ADIF), 28230 Madrid, Spain

Departamento de Física de Materiales, Facultad de Físicas, Universidad Complutense de Madrid (UCM), 28040 Madrid, Spain

Jose Maria Lopez-Villegas

Departament d'Enginyeria Electrònica i Biomèdica, Universitat de Barcelona, 08028 Barcelona, Spain

Eugene M. Chudnovsky

Graduate Program in Physics and Initiative for the Theoretical Sciences, Graduate Center, The City University of New York, New York, NY 10016, USA

Department of Physics and Astronomy, Herbert H. Lehman College, The City University of New York, Bronx, NY 10468-1589, USA

Microwave power absorption by a two-layer system deposited on a metallic surface is studied in the experimental setup emulating the response to a radar signal. Layers containing hexaferrite and iron powder in a dried paint of thickness under 1mm are used. The data is analyzed within a theoretical model derived for a bilayer system from the transmission line theory. Good agreement between experimental and theoretical results is found. The advantage of using a bilayer system over a single-layer system is demonstrated. We show how the maximum microwave absorption (minimum reflection) can be achieved through the optimization of the filling factors and thicknesses of the two layers.

**1. Introduction**

The absorption of microwaves by thin layers of composite materials with specially designed magnetic and dielectric properties has recently moved to the forefront of electromagnetic research by many physics, chemistry, and engineering labs, see, e.g., References [1,2,3] and references therein. It has been largely driven by the military applications related to radar and stealth technology, as well as by the needs of medical, educational, and research facilities to shield rooms and equipment from the background microwave radiation that has been steadily growing in the last decades due to the increase in wireless communications.

Magnetic systems have been among the most utilized materials for that purpose [4]-[7] because ferromagnetic resonance naturally falls into the microwave frequency range. The research in this area has been focused on two aspects of microwave absorption. The first is related to the absorption properties of the material per se, and how effective it is in converting microwave power into heat. To provide a significant effect, the material must be dielectric. Otherwise, the absorption would be limited to the skin layer. If the material is comprised of metallic particles, they must be either coated with an insulating layer or embedded in the dielectric matrix to suppress the overall conductivity which would result in a significant reflection of the microwaves.

The second aspect of microwave absorption, which is important for applications, arises when thin layers of the absorbing material are used. It is related to the interference of microwaves reflected by the two surfaces; the most inner surface being typically interfaced with a metal. In this case, the maximum absorption by the layer should occur at the wavelength that provides destructive interference of the two reflected waves. This effect, however, has limited use because it is sharply peaked at a certain wavelength, which stealth technology and other applications of microwave absorbers cannot rely upon. To have practical importance, the absorber must be broadband, see, e.g., References [8,9]. Our focus is on microwave absorption by magnetic bilayers. While single-layer systems have been intensively studied, with hundreds of articles entering the literature annually, experiments on bilayer systems and their theoretical analysis have been scarce.

In Reference [10] microwave absorption by nanocrystalline NiZn ferrite and iron microfibers forming single and double layers were studied in the frequency range of 2-18 GHz. The advantage of double layers in obtaining strong broadband absorption was demonstrated experimentally. A formula for the impedance of the double layer was used to fit the experimental data. Significant improvement in microwave absorption was observed. The enhancement of the absorption in the 2-18 GHz range by a double layer based upon nickel oxide and CoNiZnFeO ferrite composites, as well as the analysis of the results with the use of the impedance formula for the double layer, were reported in Reference [11]. In Reference [12] the improvement of microwave absorption properties in the 8-18 GHz range was demonstrated by using up to a 3mm double layer of carbon black/epoxy resin and NiZnFeO/epoxy resin. Microwave absorption by BaFeO and BaCoZnFeO multi-nanolayers in the 7-13 GHz range was studied and analyzed with the use of the multilayer impedance formula in Reference [13]. The optimal absorption was achieved below 500 nm total thickness. The problem of thickness optimization of a double-layered microwave absorber containing magnetite particles was investigated in Reference [14]. The procedure provided a significant increase in the reflection loss in the 8-12 GHz frequency range.

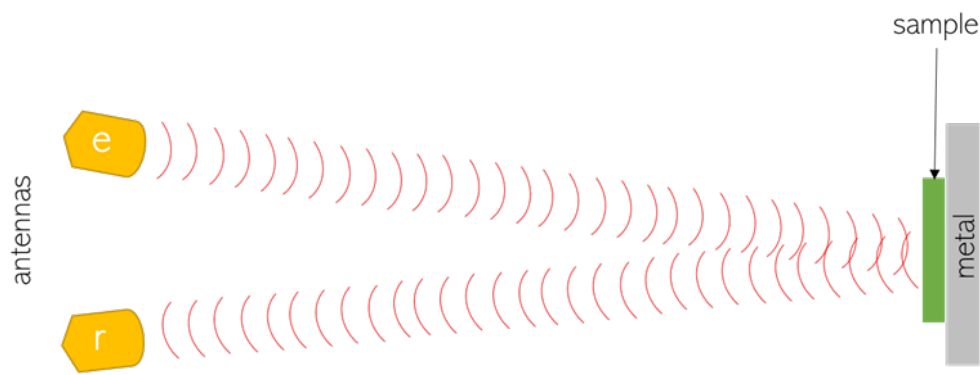

**Figure 1.** Schematic representation of the anechoic chamber measurement method in our experiments. GHz radiation is generated by the emitter (e) before interacting with the sample in a reflection mode and being detected by the receiver (r).

Here we report our effort to optimize both aspects of microwave absorption (by the material itself and due to the destructive interference) mentioned above by experimenting with bilayers based upon magnetic particles embedded in a dielectric medium. All previously reported results were obtained by studying electric and magnetic response to microwaves of small samples using a network analyzer, typically by means of waveguides or coaxial probes. The data were then plugged into the theoretical formula for the reflection loss by a layer and the prediction was made regarding the power absorption by a thin layer of the material. While the use of Maxwell equations for the absorbing medium is always justified, real systems always have features unaccounted for, such as, e.g., fluctuations in the composition and thickness of the absorbing layer. To account for such features, we directly measure the microwave response of a thin layer or bilayer of the absorbing material deposited on a macroscopic-size metallic surface, see **Figure 1**. Such a method emulates real situations relevant to stealth and shielding technology.

The paper is structured as follows. Materials and experimental methods are described at the end of the manuscript, in Section 5. The general expression for the reflection loss by a bilayer system, that we were unable to find in textbooks or published literature, is derived in Section 2. Our experimental results and their fit by theoretical formulas are presented in Section 3. Section 4 contains a discussion of the results and suggestions for applications of microwave absorbers.

## 2. Bi-layer system absorption model

The Reflection-Loss of a single layer material has been successfully analyzed using *Transmission Line Theory*: [12,13]

$$Z = Z_0 \sqrt{\frac{\mu_r}{\varepsilon_r}} \tanh\left[\left(j\frac{2\pi f d}{c}\right)\sqrt{\mu_r \varepsilon_r}\right] \qquad (1)$$

$$R_L(dB) = 20 \log \left|\frac{Z/Z_0 - 1}{Z/Z_0 + 1}\right| \qquad (2)$$

where $f$ is the frequency of the microwave, $c$ is the velocity of light in vacuum and $d$ is the thickness of the material. In this section we want to find a similar expression for a bi-layer system. In the transmission line theory, the impedance of a material can be modelled as a series of resistors, inductors and capacitors which can be related to relative permeabilities and permitivities.[12]

Following the derivation for a single layer impedance,

$$\begin{cases} \frac{\partial v(x,t)}{\partial t} = -L(x)\frac{\partial i(x,t)}{\partial x} \\ \frac{\partial i(x,t)}{\partial t} = -C(x)\frac{\partial v(x,t)}{\partial x} \end{cases} \qquad (3)$$

In a single layer system, the capacity ($C$) and the induction ($L$) are constant. In our case they are thickness dependent due to the fact that there are two layers. The system of equations can be solved assuming oscillatory behavior of the magnitudes $v(x,t) = V(x)\exp(\pm j\omega t)$ and $i(x,t) = I(x)\exp(\pm j\omega t)$.

Considering a bi-layer system with the first material $C_1$ and $L_1$ and the second material of thickness $d_2$ characterized by $C_2$ and $L_2$ the solutions are:

$$\begin{cases} V_1(x) = A\exp(j\omega\sqrt{L_1 C_1}x) + B\exp(-j\omega\sqrt{L_1 C_1}x) \\ I_1(x) = \sqrt{\frac{C_1}{L_1}}\left(-A\exp(j\omega\sqrt{L_1 C_1}x) + B\exp(-j\omega\sqrt{L_1 C_1}x)\right) \\ V(x) = E\exp(j\omega\sqrt{L_2 C_2}(x-d_1)) + F\exp(-j\omega\sqrt{L_2 C_2}(x-d_1)) \\ I_2(x) = \sqrt{\frac{C_2}{L_2}}\left(-E\exp(j\omega\sqrt{L_2 C_2}(x-d_1)) + F\exp(-j\omega\sqrt{L_2 C_2}(x-d_1))\right) \end{cases} \quad (4)$$

Imposing continuity conditions for the voltage, $V_1(d_1) = V_2(d_1)$, and for the current, $I_1(d_1) = I_2(d_1)$, yields the general expression:

$$Z = \sqrt{\frac{L_2}{C_2}} \frac{Z_M(1-\eta\tan(\omega\gamma_1 d_1)\tan(\omega\gamma_2(x-d_1))) + Z_L(j\tan(\omega\gamma_1 d_1) + j\tan(\omega\gamma_2(x-d_1)))}{Z_M(-\eta j\tan(\omega\gamma_1 d_1) - j\tan(\omega\gamma_2(x-d_1))) + Z_L(\tan(\omega\gamma_1 d_1)\tan(\omega\gamma_2(x-d_1)) - \eta)} \quad (5)$$

where $\eta = \sqrt{C_1 L_2 / L_1 C_2}$ and $\gamma_i = \sqrt{C_i L_i}$. The quantities $Z_M$ and $Z_L$ are the impedances of the material and the transmission line respectively and $x > d_1$.

The material impedance is set to zero. In $\gamma_i = \frac{\sqrt{\epsilon_i \mu_i}}{c}$, where we have dropped the subindex $r$, $\epsilon_i$ and $\mu_i$ are relative permeability and permittivity respectively, and $\sqrt{L_i/C_i} = \sqrt{\mu_i/\epsilon_i}Z_0$. We obtain the following expression (that reduces to the single layer impedance if the same material for both layers is considered):

$$Z = -Z_0 \frac{\sqrt{\frac{\mu_1}{\epsilon_1}}\tanh\left(j\frac{2\pi\omega}{c}\sqrt{\epsilon_1\mu_1}d_1\right) + \sqrt{\frac{\mu_2}{\epsilon_2}}\tanh\left(j\frac{2\pi\omega}{c}\sqrt{\epsilon_2\mu_2}d_2\right)}{1 + \sqrt{\frac{\mu_1 \epsilon_2}{\epsilon_1 \mu_2}}\tanh\left(j\frac{2\pi\omega}{c}\sqrt{\epsilon_1\mu_1}d_1\right)\tanh\left(j\frac{2\pi\omega}{c}\sqrt{\epsilon_2\mu_2}d_2\right)} \quad (6)$$

## 3. Results

The synthesized hexaferrites were first chemically and crystallographically characterized by the XRD. **Figure 2** shows the diffraction patterns obtained for each sample. The analysis of these patterns confirms the expected composition. No impurities or further structures were detected.

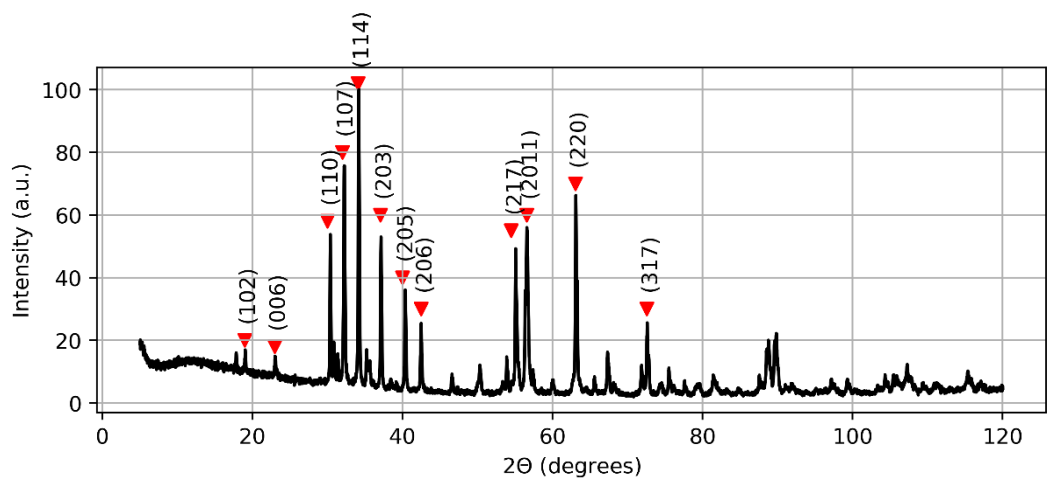

**Figure 2.** Experimental XRD pattern measured for *HF* powder sample. The most significant peaks are identified with the index of the plane corresponding to each reflection.

The *HF* and *Fe* powder samples, together with the commercial paint, were all electromagnetically characterized in the GHz frequency range with the coaxial probe. The Nicolson-Ross-Weir (NRW) model [12]–[14] was used to deduce the complex magnetic permeability and dielectric permittivity from the measured complex *S*-parameters. In the case of the paint, which is a non-magnetic material, the NRW method has been adapted to pure dielectric conditions.[15] **Figure 3** depicts the complex spectra obtained for each of the three materials, both for the electric and magnetic contributions.

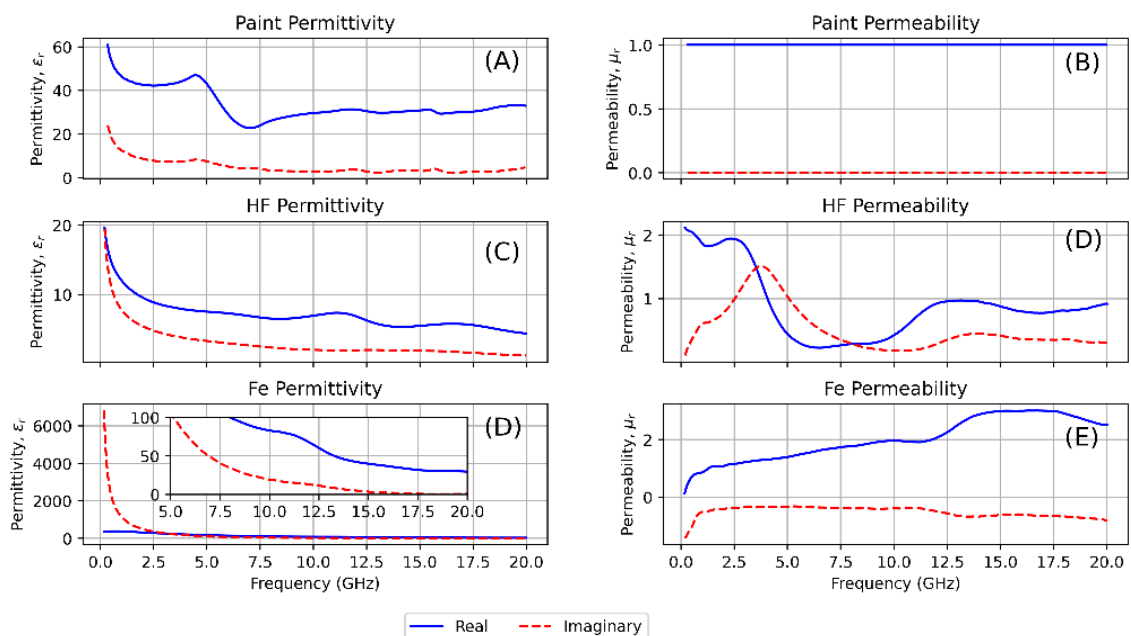

**Figure 3.** Complex permittivity (left column) and permeability (right column) for the paint (A and B), the hexaferrite (C and D) and the iron powder (E, and F). Data obtained after processing the *S*-parameter through the NWR method.

As it may be observed, the Fe permeability (Figure 3 (E)) is lower than the typical one at the kHz and MHz frequency range because the magnetic losses (and the consequent drop in permeability) appear at a few MHz [1]. Moreover, measuring the powder samples instead of bulk ones may increase the demagnetization field effects, which leads to a decrease of the measured permittivity as well. Nonetheless, this is still the factor that provides the highest magnetic permeability in the system. Besides, the metallic particles' permittivity is also large (Figure 3 (D)) given that their free electrons may easily be displaced by the electric field, leading to a strong polarization.

The effective electromagnetic properties of the sheets (which contain a mixture of paint and functional powder) have been modeled by using the Maxwell-Garnett (MG) model:[16], [17]

$$\frac{\varepsilon_{eff}-\varepsilon_h}{\varepsilon_{eff}+2\varepsilon_h} = ff_v \frac{\varepsilon_i-\varepsilon_h}{\varepsilon_i+2\varepsilon_h} \qquad (7)$$

where $\varepsilon_{eff}$ refers to the effective (paint + powder) permittivity, while $\varepsilon_h$ and $\varepsilon_i$ refer to the host (paint) and inclusions (powder) permittivity. The same conversion has been done with the permeability. Given that this model depends on the volume filling factor ($ff_V$), reference densities of 1.50 g/cm³, 5.30 g/cm³ and 7.87 g/cm³ have been used for the paint, ceramic and metallic components – respectively – for converting the weight fill factor ($ff_W$) to the volumetric one ($ff_V$). The resulting effective electromagnetic properties are, obviously, dependent on the filling factor. Given that our sheets are not uniform and therefore we do have a concentration gradient in the radial direction, we have run this computation for different $ff_V$ (i.e., $ff_W$) to produce a clear picture of the electromagnetic properties of the samples. This modeling is represented in **Figure 4**, which shows the complex permittivity and permeability obtained for each mixture of materials and $ff_W$ of 4%, 20%, 40%, 60% and 80%, which correspond to $ff_V$ of 1.17%, 6.61%, 15.87%, 29.80% and 53.01% for samples containing *HF* (P1, P2 and P3), and $ff_V$ of 0.79%, 4.55%, 11.27%, 22.23% and 43.26% for those with *Fe* powder (P4, P5 and P6).

As it may be observed, for low filling factors both components have very similar electromagnetic properties compared to the pure paint, which is the major component. While the filling factor increases, the effective properties tend to be closer to the ones of the functional particles dispersed in the paint and each type of layer starts to have a more particular behavior. As it has been mentioned before, our sheets have a global $ff_W$ = 4% but have a concentration gradient from the center to the edges. Therefore, at the center of the sheet $ff_W$ > 4%, where the interaction with the electromagnetic radiation is stronger. According to our estimations, the sheets have $ff_w \sim 11\%$ at the 15×15 cm² central region, $ff_w \sim 25\%$ at the central 10×10 cm² or even $ff_w \sim 70\%$ at the 6×6 cm² central part. Therefore, we must consider later with our model the possibility of having these considerable amounts of loading at the region with the strongest electromagnetic interaction.

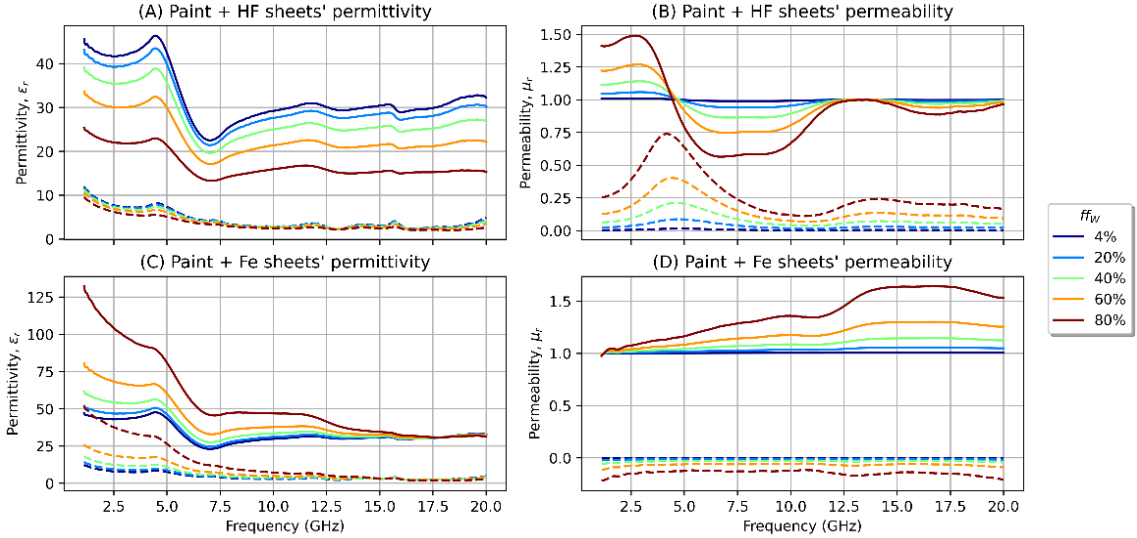

**Figure 4.** Modeling of the mixture's effective properties by applying the MG model to the permittivity and permeability data of each of the components using $ff_w$ of 4%, 20%, 40%, 60% and 80%.

With these permittivity and permeability complex data, we first validate the $R_L$ results obtained for each individual layer. To do so, we compare the experimentally measured $R_L$ at the anechoic chamber with the $R_L$ calculated by the *Transmission Line Theory* using the sheets' electromagnetic properties measured before. This comparison is represented in **Figure 5**, being the first row of subplots devoted to sheets P1-P3 (paint + *HF*) and the second one to P4-P6 (paint + *Fe*). In each case, first the comparison between the experimental and calculated values is done for the three thicknesses prepared at the laboratory. On the right side, a 2D $R_L$ simulation for a wider and continuous range of thicknesses is done in order to have a broad view of the expected absorption of each sheet, even outside of the experimental domain. The agreement between the experimental and the calculated $R_L$ data has been verified from $ff_w = 4\%$ to 80%.

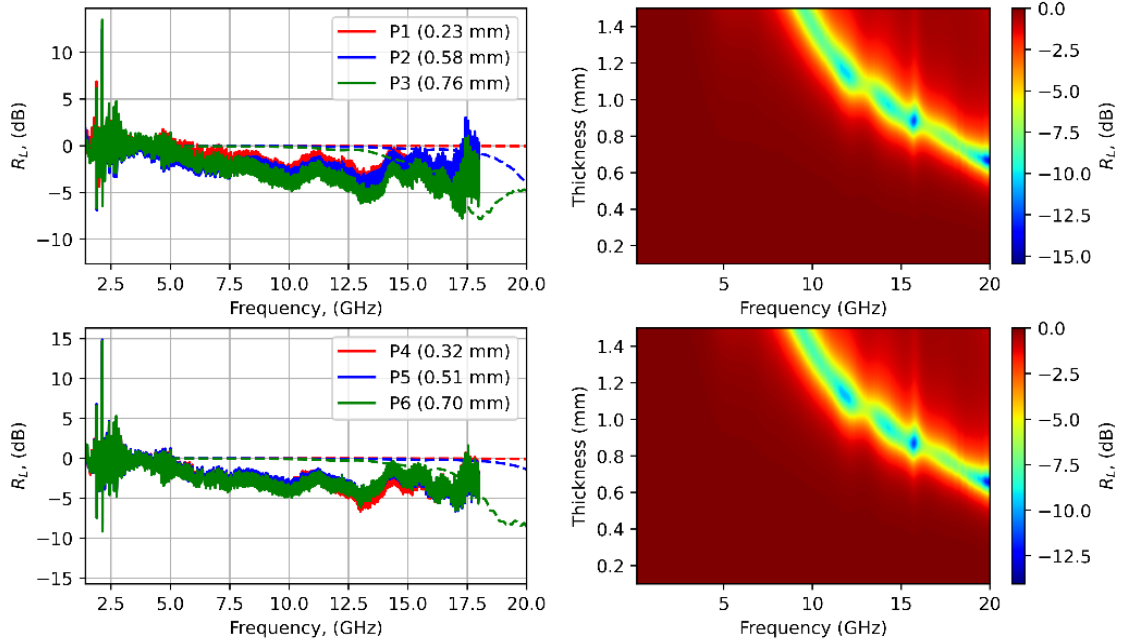

**Figure 5.** Comparison between the experimental and calculated data for the sheets containing *ff_W* = 4% of *HF* (top row) and *Fe* (bottom row) particles. Left side plots correspond to the comparison for the specific thicknesses prepared at the laboratory, while on the right side there is a 2D simulation of the expected $R_L$ for sheets between 0.1 and 1.5 mm. Solid lines correspond to experimental data, while dashed lines correspond to the calculated values.

One may immediately see that none of the layers presents significant electromagnetic absorption when being irradiated with microwaves between 0.5 and 18 GHz. The spectra are flat and the small deviation from the zero is attributed to experimental noise. The calculations done with the single-layer *Transmission Line Theory* reinforce these observations, as no absorption is expected in this frequency range for any of the compositions used.

The 2D simulations provide some additional information of interest. According to these predictions, we should not observe any absorption peaks for samples thinner than 0.8 mm below 18 GHz, as happened in the real radar experiments. Moreover, these results suggest that slightly thicker samples would start to absorb radiation in this frequency range. This prediction agrees with the experimental data, which seems to show the start of a peak at the top frequency limit for the thicker sample, for both types of samples. These simulations also predict that absorptions up to 12.5 and 15.0 dB would be feasible for the *HF* and *Fe* sheets, respectively. Given that both types of sheets have a predominant paint content (and thus, their electromagnetic properties do not diverge that much) the resultant $R_L$ spectra are also similar, though small changes are appreciated due to the functional powder effect.

Once that single layer systems have been discussed, we move to double layer systems, which represent a much more unexplored territory in literature, both theoretically and experimentally. To do so, again we provide a detailed comparison between the anechoic chamber measured data and the predictions done by our own model using the measured complex electromagnetic properties. Let's start by analyzing a system consisting of the combination of the thickest *HF* layer (P3) with the different thickness *Fe* sheets, i.e., we have a system of two layer with different electromagnetic properties and varying thickness. **Figure 6** below shows the experimental data, measured in the anechoic chamber, together with the calculation from the bilayer model for $ff_w$ = 20%, 40%, 60% and 80%.

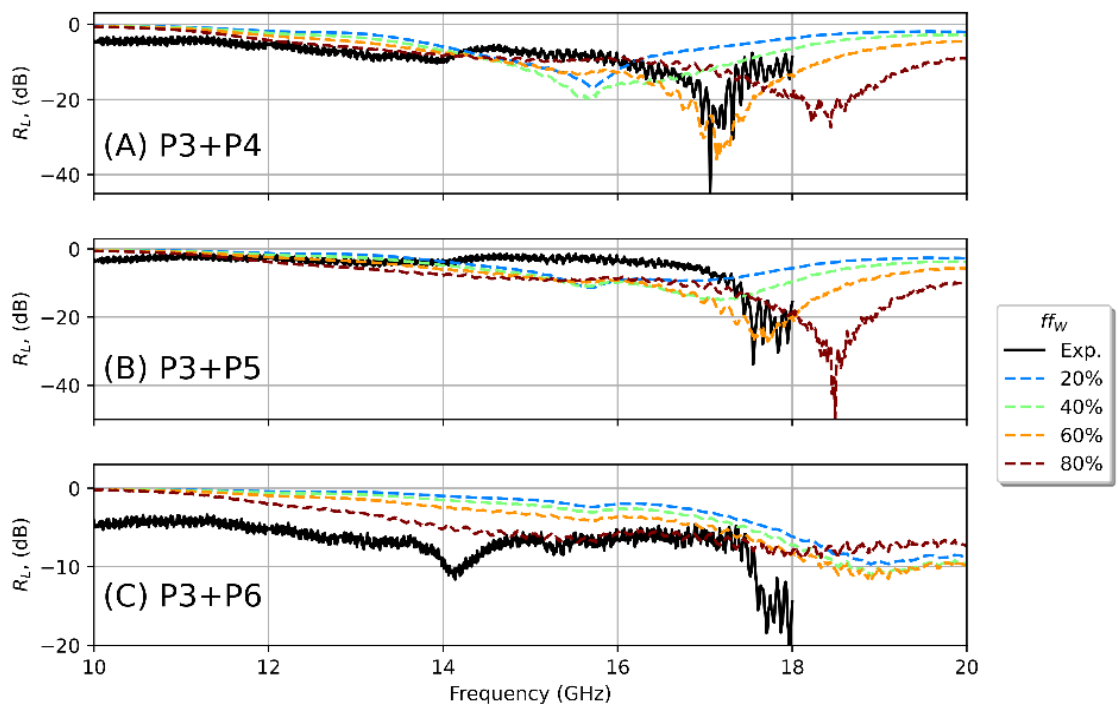

**Figure 6.** Comparison between the experimental (solid black) $R_L$ measured in the anechoic chamber and the $R_L$ data calculated by the model (dashed) with $ff_w$ = 20%, 40%, 60% and 80%, for the bilayered systems (A) P3+ P4, (B) P3+P5 and (C) P3+P6.

As a first observation, it must be highlighted that that the model is capable, in all cases, to reproduce with reasonable accuracy the peak position, amplitude and shape from the thickness and complex permittivity and permeability data. Looking in more detail, we may observe that – as is expected from the gradient in particles' concentration – the agreement is not so good for low filling factors. However, when the filling factor is increased, the theoretical and the experimental data match. The best agreement is reached for $ff_w$ = 60%, giving us an idea about the real filling factor at the center of the sheet. If the $ff_w$ becomes too high (80%), the agreement deteriorates. For cases (A) P3+P4 and (B) P3+P5 the agreement is quite extraordinary, while in (C) P3+P6 the model tends to predict the peak that seems to move to the frequency range above 18 GHz.

From Figure 6(A) it is possible to extract a few interesting conclusions. As one may observe, if the filling factor is too high (80%) the agreement worsens and the amplitude of the calculated peak is reduced. This highlights the importance of optimizing and selecting the appropriate filling factor in the material design process and during the computational simulation of $R_L$. In the first case, this is a crucial aspect, just increasing the magnetic powder load does not always result in the increase of the power absorption. It is necessary to optimize the filling factors based on the electromagnetic properties of each component, to reach the adequate effective permittivity and permeability that create an impedance match between the two layers of specific thickness. That is a crucial point because, from the modelling perspective, the reflection loss has a *tanh* dependance on the permittivity, permeability and the thicknesses of the two layers, so it is very sensitive to small changes in these variables. Such an optimization within our theoretical model is shown in **Figure 7**.

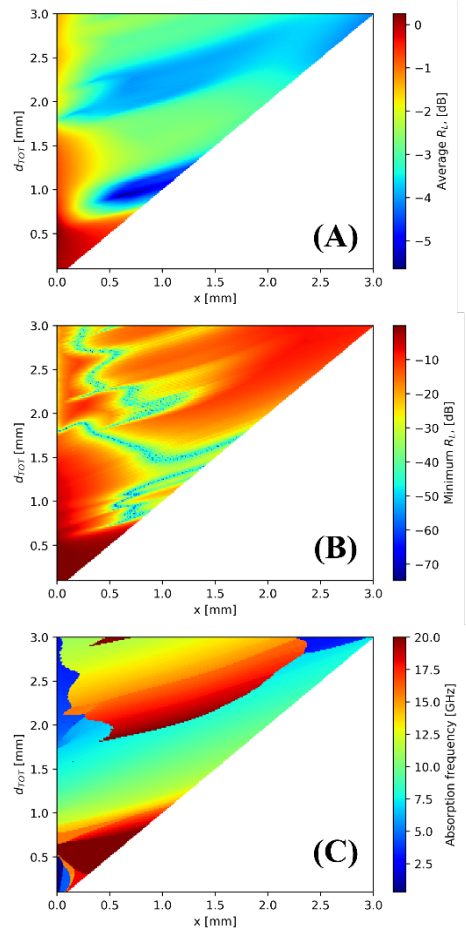

**Figure 7.** 2D simulation of the (A) average $R_L$, (B) minimum $R_L$ (maximum absorption) and (C) absorption frequency as a function of the thickness of each layer in a bi-layer system. $x$ refers to the thickness of the first layer, while $d_{TOT}$ is the total thickness of the bilayer.

The simulations shown in the figure above are a powerful tool for design processes of high absorption materials. In the first case, Figure 7(A), represents the average reflection loss produced by the sample all along the analyzed frequency range. Therefore, it is affected by the peak base but not only by the maximum peak amplitude, leading to logic smaller $R_L$. Nonetheless, this figure is of great importance because it shows an overall absorption behavior of the material along the frequency range, considering both the peak amplitude and width. On the other hand, Figure 7(B) represents the maximum absorption (minimum $R_L$) achieved for each combination of thicknesses between the two layers. As it may be observed, losses around 40 – 50 dBs may be easily achieved with a first layer of thickness between 0.5 and 1 mm and below 0.5 mm in the second one. As stated before, the constant increase of thickness does not lead to a permanent increase in absorbed power. Thus, optimizing the relative thickness between layers is a crucial factor do design materials with the best possible shielding capabilities. Finally, Figure 7(C) provides a clear view of the dependance of the resonance (and, thus, absorption) frequency on the thicknesses of the two layers. One may find discontinuities in this spectrum (i.e., abrupt changes in frequency), which may either be due to the noise contribution when the peak's signal-to-noise ratio is low or because in some cases there may be several absorption peaks with one stronger than the others.

## 4. Conclusion

We have studied microwave absorption by a two-layer system containing powders of iron and barium hexaferrite particles. Layers under 1-mm thickness of dried paint containing the particles were deposited on a metallic surface. The reflection loss has been studied in an experiment emulating response to a radar signal. This differs from the experiments reported in literature, where the reflection loss was not directly measured but calculated from the data on frequency dependence of real and imaginary parts of permittivity and permeability. While such experiments are valuable, they ignore many features present in the reflection of a radar signal, such as the geometry of the reflecting system inhomogeneity of the parameters.

We began with establishing that the material of paint, which dominated composition of the layers, was not responsible for any reflection loss. We then studied microwave absorption by single layers of different thickness and filling factors. The measured reflection loss from single layers was compared with a theoretical formula and a good agreement was found. The magnitude of the reflection loss for single layers was low, however, at frequencies up to 18GHz in all ranges of the parameters used. The observed absorption changes substantially with any changes in the effective permittivity or permeability of any of the two layers. This occurs on either changing the functional filler powders or by changing the filling factor. Any change in the electromagnetic properties of the layers requires a new optimization of their thicknesses to maximize the absorption.

This changed dramatically when we moved to a bilayer system with the same parameter ranges for single layers. The reflection loss in some cases has grown ten-fold, while still allowing the total thickness of the bilayer system under 1mm. The data were analyzed with the use of a theoretical expression for a bilayer system derived by methods of the transmission line theory. Good agreement between experimental and theoretical results has been found. The ratio of the thicknesses of the two layers as an additional optimization parameter in the impedance match has been shown to play a pivotal role in the enhancement of the reflection loss.

## 5. Experimental Section/Methods

The preparation of the absorbing samples requires multiple components that need to be carefully processed in order to produce sheets with the desired properties. The most significant element in this preparation is the functional powder, which is the material that provides the specific dielectric and magnetic character to the sample. We worked with two types of powder materials: barium-hexaferrite ceramic materials (from now, noted as "*HF*") and soft magnetic iron particles (referred as "*Fe*"). The *Fe* powders were provided by AMES enterprise and a complete structural and magnetic characterization may be found in our previous works.[1], [2] It consists of Fe-core particles of 100 μm in diameter with high permeability, which are widely used for kHz and MHz applications. The core's crystallinity is high, and no impurities are detected by XRD. On the other hand, the *HF* powder samples were synthesized by our own at the laboratory, and have been selected for this work because they have extended reputation as microwave absorbers.[8]–[11]

*Synthesis of functional powder*

The barium hexaferrite ($BaFe_{12}O_{19}$) powder was synthesized through conventional co-precipitation route. Stoichiometric amounts of $Ba(NO_3)_2$, $Fe(NO_3)_3 \cdot 9H_2O$ (Scharlab, as received) were dissolved in 300 mL of deionized water under continuous stirring at room temperature for two hours. Once complete dissolution was reached, a 2M solution of NaOH was added dropwise until reaching pH ~ 10. The brown precipitate was cleaned and separated

13with any changes in the effective permittivity or permeability of any of the two layers. This occurs on either changing the functional filler powders or by changing the filling factor. Any change in the electromagnetic properties of the layers requires a new optimization of their thicknesses to maximize the absorption.

This changed dramatically when we moved to a bilayer system with the same parameter ranges for single layers. The reflection loss in some cases has grown ten-fold, while still allowing the total thickness of the bilayer system under 1mm. The data were analyzed with the use of a theoretical expression for a bilayer system derived by methods of the transmission line theory. Good agreement between experimental and theoretical results has been found. The ratio of the thicknesses of the two layers as an additional optimization parameter in the impedance match has been shown to play a pivotal role in the enhancement of the reflection loss.

## 5. Experimental Section/Methods

The preparation of the absorbing samples requires multiple components that need to be carefully processed in order to produce sheets with the desired properties. The most significant element in this preparation is the functional powder, which is the material that provides the specific dielectric and magnetic character to the sample. We worked with two types of powder materials: barium-hexaferrite ceramic materials (from now, noted as "*HF*") and soft magnetic iron particles (referred as "*Fe*"). The *Fe* powders were provided by AMES enterprise and a complete structural and magnetic characterization may be found in our previous works.[1], [2] It consists of Fe-core particles of 100 μm in diameter with high permeability, which are widely used for kHz and MHz applications. The core's crystallinity is high, and no impurities are detected by XRD. On the other hand, the *HF* powder samples were synthesized by our own at the laboratory, and have been selected for this work because they have extended reputation as microwave absorbers.[8]–[11]

*Synthesis of functional powder*

The barium hexaferrite ($BaFe_{12}O_{19}$) powder was synthesized through conventional co-precipitation route. Stoichiometric amounts of $Ba(NO_3)_2$, $Fe(NO_3)_3 \cdot 9H_2O$ (Scharlab, as received) were dissolved in 300 mL of deionized water under continuous stirring at room temperature for two hours. Once complete dissolution was reached, a 2M solution of NaOH was added dropwise until reaching pH ~ 10. The brown precipitate was cleaned and separated from water by doing 3 cycles of 10 minutes of centrifugation at 3000 rpm with ethanol. The obtained solid part was dried 24 hours at 80°C and then ground to powder with a mortar before being heated at 900°C for 2 hours in a furnace. The final product was finally grounded again.

*Sheets preparation*

The functional powders need to be later processed in order to produce two-dimensional sheets, in such a way that the material can be deposited into a surface. To do so, the functional powder was mixed with paint (commercial Titan's Unilak white water enamel) in a 4%-96% weight fraction ($f_W$), respectively. The mixture was then manually deposited on top of a 50 μm thickness polyester sheet of 25 cm × 25 cm. The painted sheets were left drying at room temperature for 24 hours.

For each type of functional powder, we prepared three different sheets varying the total thickness, as described in Table 1 below. Therefore, we are not only able to evaluate the effect of the intrinsic electromagnetic properties of the powder dispersed on it, but also to figure out the repercussion that a geometrical aspect has on the absorption procedure.

**Table 1.** Description of the prepared sheets. Thickness values are obtained as the average of four measurements at the sheet's contour.

| Sheet ID | Contained powder ID | Total thickness (mm) |
|---|---|---|
| P1 | *HF* | 0.23 ± 0.07 |
| P2 | *HF* | 0.58 ± 0.05 |
| P3 | *HF* | 0.76 ± 0.09 |
| P4 | *Fe* | 0.32 ± 0.09 |
| P5 | *Fe* | 0.51 ± 0.08 |
| P6 | *Fe* | 0.70 ± 0.06 |

Due to the manual deposition method used for the sheets' preparation and the elevated density of the magnetic powder, the obtained sheets do not show a homogeneous dispersion of magnetic particles along its surface. Consequently, all sheets show a larger concentration of particles at their central region compared to the perimeter, leading to an effective gradient of filling factor from the center to the borders.

*Characterization*

Powder x-ray diffraction (XRD) measurements were done by using a PANalytical X'Pert PRO MPD θ/θ Bragg−Brentano powder diffractometer of 240 mm radius using Cu *K-alpha* radiation ($\lambda$ = 1.5418 Å). The static magnetic properties from powder samples were measuring through a Quantum Design MPMS Superconducting QUantum Interference Device (SQUID) magnetometer. On the other hand, in order to measure the complex magnetic permeability and dielectric permittivity in the microwave region (from 0.05 to 20 GHz) we designed a coaxial probe by using two coaxial 3.5 mm connectors coupled to a Keysight E5071C ENA Series Network Analyzer. The system was electronically calibrated prior to the measurements and two-port *S* parameters were recorded along 1601 points from 0.05 to 20 GHz. Port-extension correction was also done to all the measurements to deal with all the potential signal phase shifts due to the probe dimensions. Finally, the experimental absorption of the sheets was measured inside an anechoic chamber. The sample was subjected on a metallic plate while two antennas were oriented in reflection configuration (see Figure 1). An Agilent E8362B PNA Series Network Analyzer was connected to the antennas and the reflection loss was recorded along 1601 points from 0.5 GHz to 18 GHz.

**Acknowledgements**

The work at the University of Barcelona has been supported by the U.S. Air Force Office of Scientific Research (AFOSR) through grant No. FA8655-22-1-7049. The work at CUNY has been supported by the AFOSR through grant No. FA9550-20-1-0299. The authors also acknowledge AMES enterprise for their collaboration and for providing necessary materials.

Received: ((will be filled in by the editorial staff))
Revised: ((will be filled in by the editorial staff))
Published online: ((will be filled in by the editorial staff))